# New Cooled Feeds for the Allen Telescope Array

Wm. J. Welch[1], Matthew Fleming[2], Chris Munson[3,4], Jill Tarter[3], G.R. Harp[3], Robert Spencer[2], Niklas Wadefalk[5]

[1] Professor in the Graduate School, UCB, Berkeley CA, Emeritus Prof. Astronomy & EECS
[2] Minex Engineering Corporation, Antioch CA
[3] SETI Institute, Mountain View CA
[4] Millennium Engineering and Integration Company, Moffett Field CA
[5] Low Noise Factory, Mölndal, Sweden

## Abstract

We have developed a new generation of low noise, broadband feeds for the Allen Telescope Array at the Hat Creek Observatory in Northern California.  The new feeds operate over the frequency range 0.9 to 14 GHz.   The noise temperatures of the feeds have been substantially improved by cooling the entire feed structure as well as the low noise amplifiers to 70 K.   To achieve this improved performance, the new feeds are mounted in glass vacuum bottles with plastic lenses that maximize the microwave transmission through the bottles.  Both the cooled feeds and their low noise amplifiers produce total system temperatures that are in the range 25-30 K from 1 GHz to 5 GHz and 40-50 K up to 12.5 GHz.

## Introduction

We describe the development of new low noise feeds to upgrade the receiver systems on the Allen Telescope Array (Welch *et. al.*, 2009).   The array consists of 42 antennas distributed over a few acres of land at the Hat Creek Radio Observatory in northern California.  Each antenna utilizes a 6.1 meter diameter primary reflector and a 2.4 meter secondary with offset-Gregorian optics to minimize the interference scattered into the system. Figure (1a) is a drawing showing the placement of the wideband feed in the 2.2% of the area of the primary reflector that is shadowed by the secondary mirror, and the metallic shroud that shields the feed from thermal ground emission. Figure (1b) is a photograph of one of the antennas, showing the Sunbrella fabric radome cover that protects the feed from the weather. The ATA is currently capable of synthesizing images with wide bandwidths, modest sensitivity, and good angular resolution.  In order to achieve this bandwidth, we use log periodic antennas, members of the class of Frequency Independent Antennas (Rumsey, 1966), as the antenna feeds. The ATA feed is pyramid shaped and receives orthogonal linear polarizations from pairs of feed arms on opposite sides of the pyramid. At any wavelength, the feed's active region is located where the width of the pyramid is about one half wavelength; this is near the wide end of the feed for long wavelengths and near the narrow end for short wavelengths. For this class of antenna, signals are received from the direction of the tip of the feed, where the input terminals are located. Received signals travel up to the tip on the feed arm structure and then onto coaxial cables down to the LNA (low noise amplifier) within the feed. Figure (2) shows a photograph of the first generation  (uncooled) feed mounted in the antenna.  A cutout drawing of the feed interior shows one polarization of the LNA cooled in a dewar to 70 K.

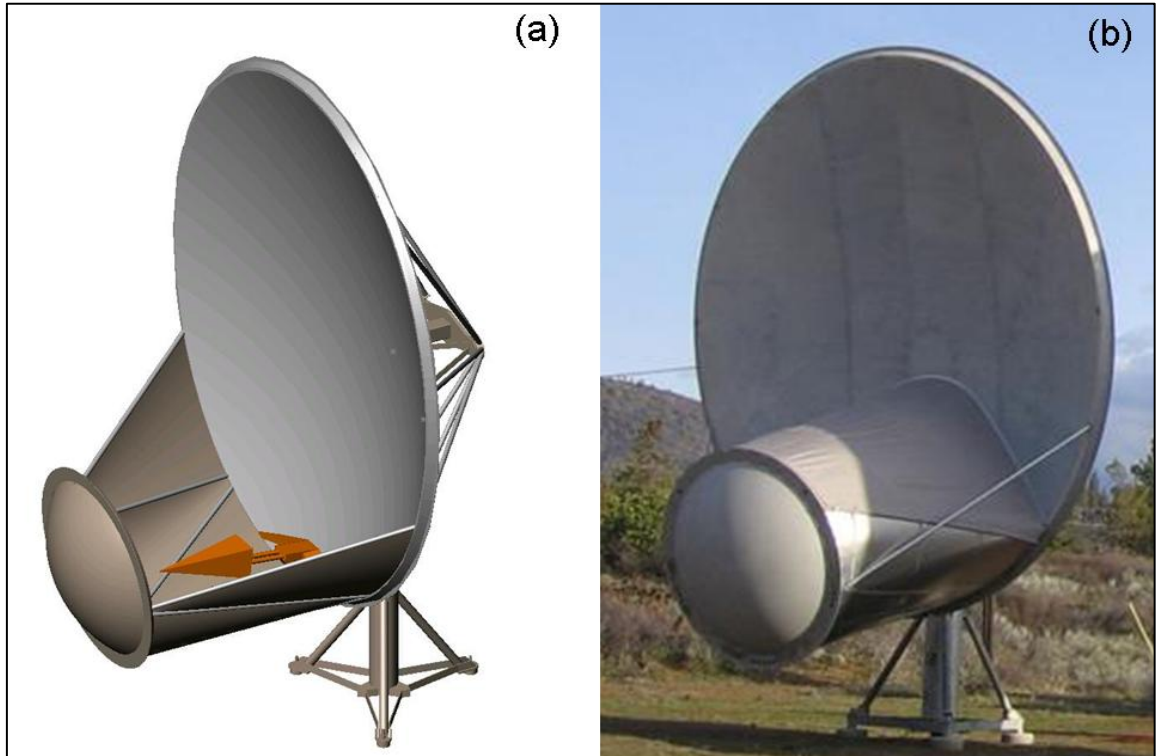

Figure (1)  (a) shows a drawing of the two-mirror offset Gregorian optics with the feed and its mounting bracket and the metal shroud between the two mirrors at the bottom.  The metal shroud protects the optical system from background radiation from the ground.    (b) is a photograph showing the Sunbrella radome cover that protects the feed from the weather.  Laboratory measurements show that this fabric has a nearly frequency independent absorption of just one percent.   Image adapted from Fig. 2 in Welch *et. al.* (2009).

An important feature of the feed and reflector system is its inherent wide bandwidth.  At any wavelength, the optimum sensitivity is achieved when the active region of the feed is placed at the fixed focus of the Gregorian optics. For observations at a single frequency, the feed is translated into position for maximum sensitivity. Figure (3) shows the variation of relative system sensitivity of the current log-period feed as a function of frequency for a number of different focus positions. The performance has a rather broad maximum; thus, a single focus position can be chosen for observations covering a wide range of frequencies. For example, if the feed is set at the 10 GHz focal point, the relative gain is > 0.9 over the range of 4.5 GHz to 15 GHz, an unusually wide bandwidth.

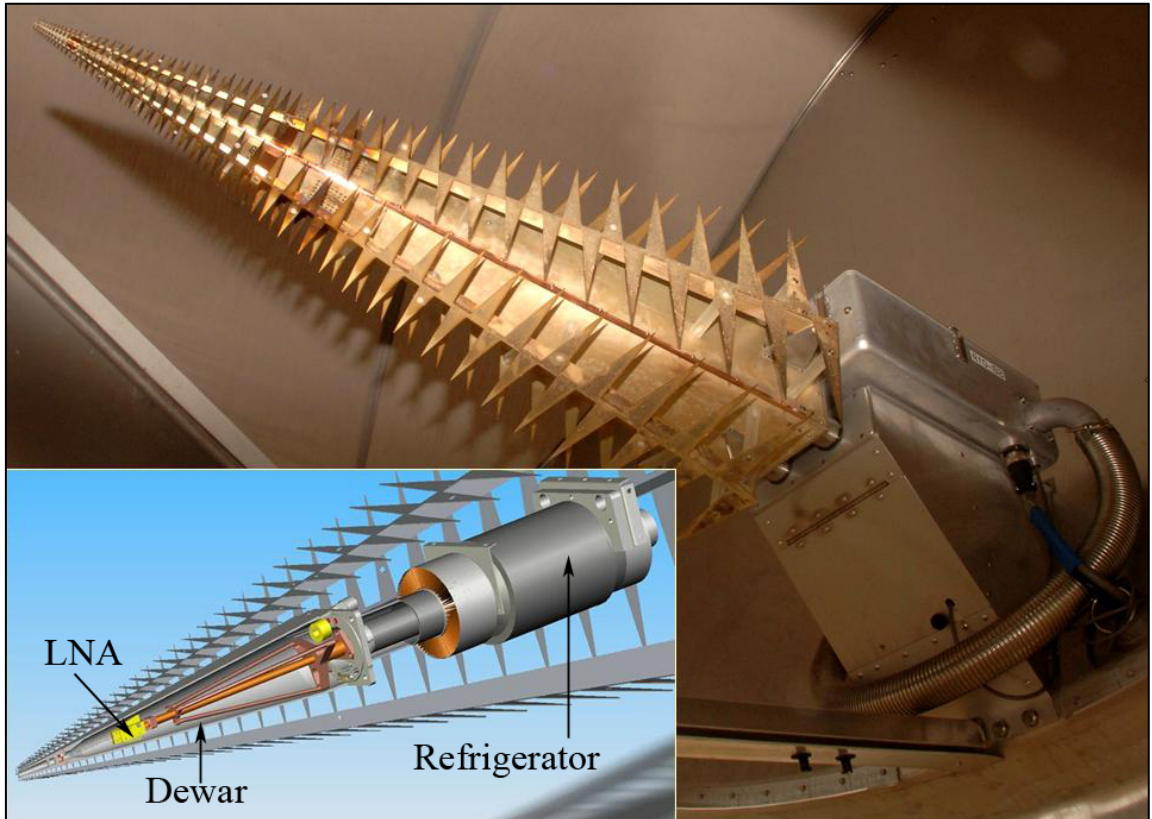

Figure (2) Current Log-Periodic Feed, with operational bandwidth from 0.5 to 10 GHz. The cryogenic dewar and LNAs are cooled to ~70 K by a commercial refrigerator mounted internal to the pyramid. Image adapted from Fig. 4 in Welch *et. al.* (2009).

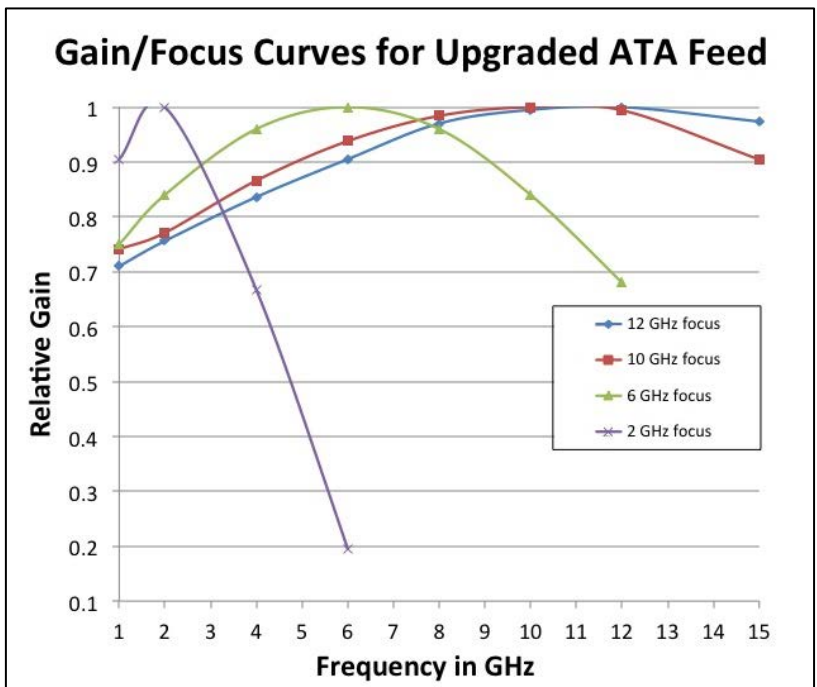

Figure (3) Change in relative sensitivity as a function of frequency for a range of different focus settings.

## The Upgrade

Welch *et. al.* (2009) described the motivations behind construction of the current wideband feeds on the ATA, and their expected performance. Since their installation, those feeds have proven to be less robust than forecast, with a number of the feeds now in need of repair and performance of some of the others degrading over time. The new feeds have been redesigned to alleviate the initial shortcomings, plus extending the highest frequency and improving noise performance at all frequencies.

A major improvement follows from the entire feed structure being cooled to a temperature close to the temperature of the LNA, 70 K. In addition, the coaxial cables between the feed tip and the LNA are also cooled to 70 K. At this lower physical temperature the resistive losses in both the feed structure and cables are much lower as is their emitted blackbody radiation. The resistivity of copper at low temperatures can be approximated as $\rho = .67 \ (T - 20/132)^{1.35} \ x10^{-6} \ \Omega$ cm. Going from 300 K to 70 K lowers the resistivity of copper from 1.85 $x10^{-6}$ to 0.18 $x10^{-6}$ $\Omega$ cm. The resistive losses scale as the surface resistance Rs, and Rs $\sim \sqrt{\rho}$. Thus the resistive losses will be a factor of 0.31 smaller. The blackbody radiation scales linearly with temperature and is thus reduced by a factor of 0.23. The product of these two temperature effects lower the received background noise at each LNA by a factor of $\sim$14. Laboratory measurements of the receiver temperature average 30 K from 1 to 10 GHz. Measurements of the LNA noise temperature contribution from the LNF-ABLNC1_15A rises from 10 to 15 K over the frequency range. Thus the contribution from the cooled feed structure and coax cables ranges from 12 to 7 K over 1 to 10 GHz.

Lowering the physical temperature of the entire feed from ambient temperature (about 300 K to 70 K), is a major challenge. Prototyping and final implementation for the Antonio Feeds was carried out at Minex Engineering in Antioch CA. The lowest frequency of the feed was moved from 0.5 to 0.9 GHz to significantly shorten its physical size. As a result, the entire feed structure can be readily contained within a vacuum vessel that is largely transparent to radio waves. This allows the cooled feed to couple well to wideband incoming signals. After testing several possible materials for the transparent dewar, we chose pyrex glass because of its very low radio frequency loss, mechanical strength, and availability, and despite its rather large dielectric constant of 5.1.

The glass vacuum bottle is a hemisphere at the tip of the feed and continues as a cone and then a cylinder past the large base of the feed, as can be seen in Figure (4). In addition, in order to improve radio transparency at the higher frequencies, we added a thin hemispherical layer of plastic, effectively a lens, with a dielectric constant of 2.3 and thickness .127 cm covering the tip of the pyrex bottle. The pyrex bottle thickness is .100 cm over the hemispheric region, with inner and outer radii at 7.620 cm and 7.720 cm respectively. The thickness of the glass grows to .500 cm at the base of the feed.
Figure (5) shows a picture of the feed antenna in the vacuum bottle with the hemispherical polyethylene lens over its tip. The optimum lens thickness and separation of the glass from the lens were found from a large number of trial calculations of the transmission of radio waves through the glass bottle and lens to the feed. For frequencies below about 5 GHz, the incoming signal goes past the edge of the plastic lens and proceeds directly to the feed through the lower part of the bottle.

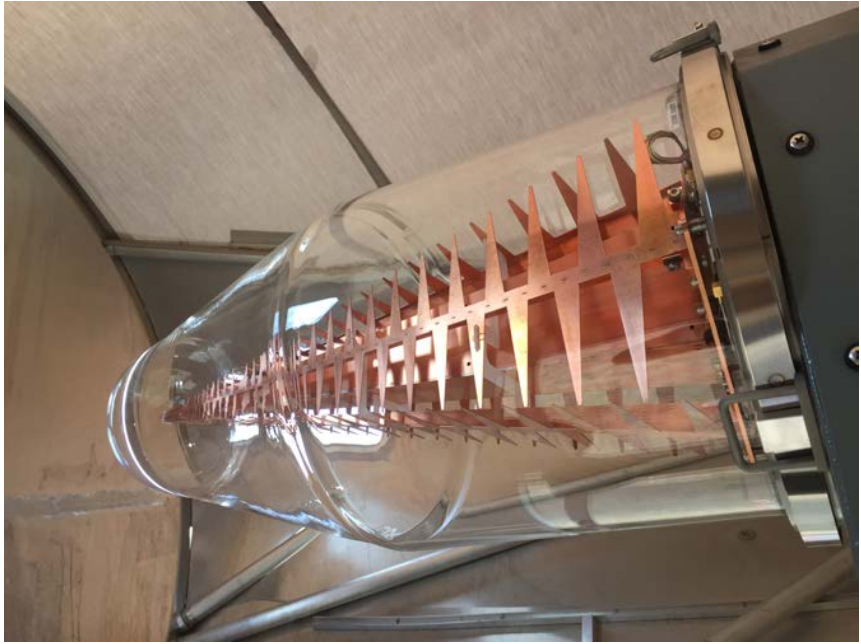

Figure (4) – Upgraded feed mounted in an ATA antenna, with geometry of the glass bottle clearly visible.

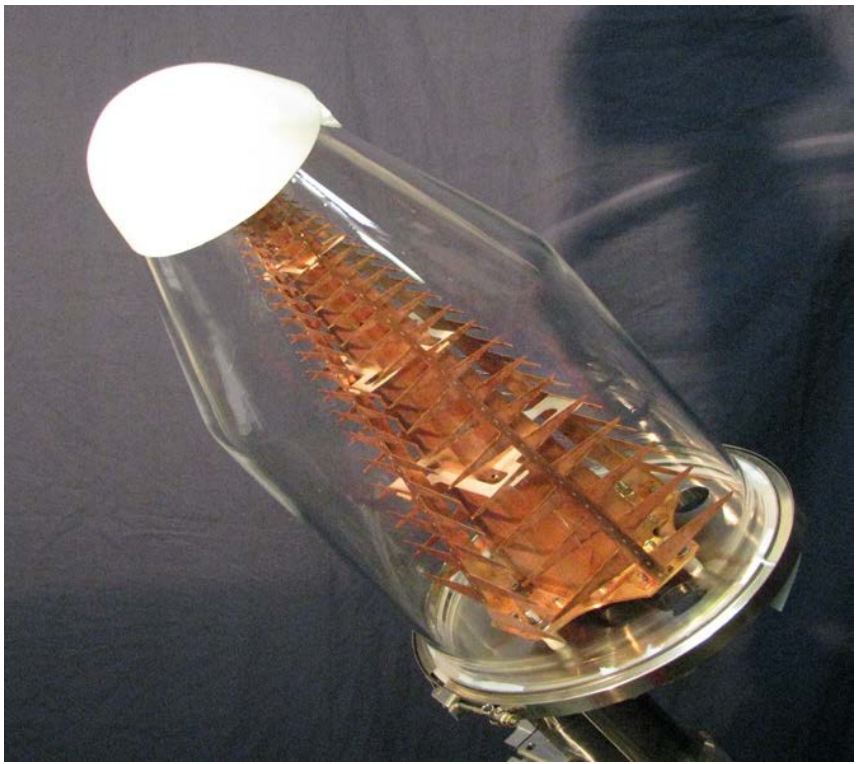

Figure (5 ) – Upgraded feed with plastic lens.

Figure (6) shows the calculated radio wavelength transmission into the feed as a function of frequency from about .9 GHz up to14 GHz for three different lens outer radii: 8.247cm, 8.267cm, and 8.287cm.  For operation up to 14GHz, the limit of the receiver, the mean radio frequency power transmission factor is .93 for the chosen radius of 8.287cm. The magenta line in Figure (6) shows the significant loss of transmission at high frequencies in the absence of any lens.

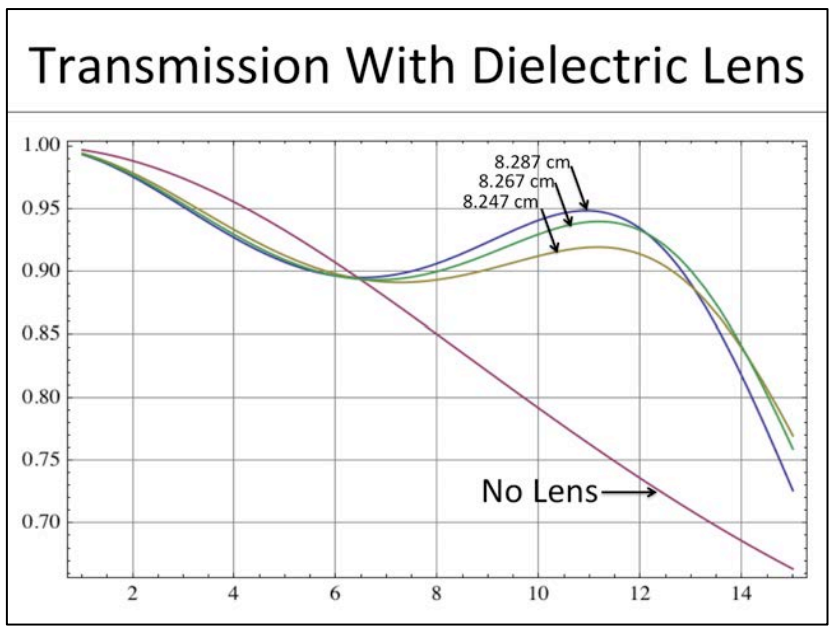

Figure (6) – Transmission through the glass and plastic lens for the three different outer radii.  The magenta curve shows the transmission in the absence of any lens.

The new Antonio feeds incorporate cryogenic amplifier MMIC chips (part number LNF-ABLNC1_15A) fabricated by Low Noise Factory as part of a research collaboration with Chalmers University and mounted on custom blocks designed to fit within the feed pyramid.  Figure (7) shows the test results demonstrating excellent gain and noise temperature for this LNA over a wide range of frequency from 0.9 – 14 GHz when cooled to cryogenic temperature.

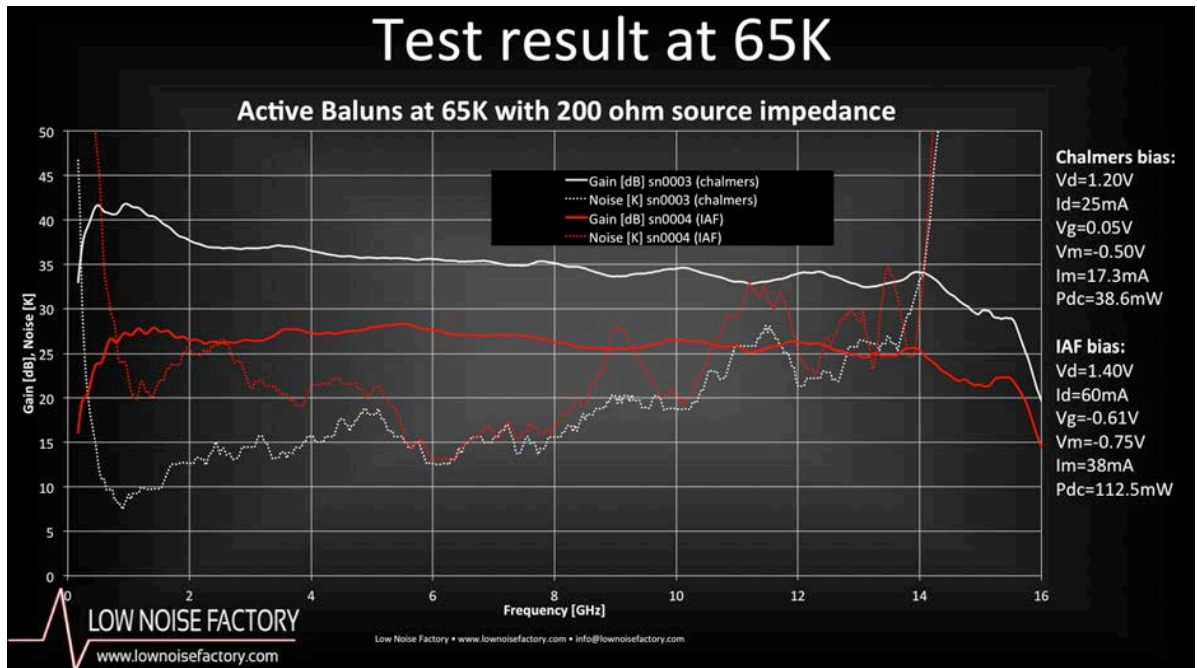

Figure (7) – The measured gain (solid white) and noise temperature in °K (dashed white) of LNF-ABLNC1_15A at cryogenic temperature. Red curves refer to an older LNA design manufactured by Fraunhofer IAF. Chart courtesy of Low Noise Factory

## Connection between the feed and the Low Noise Amplifiers

The input terminals of the two orthogonal, balanced, linearly polarized, log periodic feeds are at the four sides of the tip of the pyramidal structure. Electromagnetic transmission simulations of the original feed gave a frequency independent balanced feed impedance of 207 ohms for each linearly polarized feed (Engargiola and Welch, 2004). We use two pairs of small diameter coaxial 90 Ohm cables to connect the tip terminals to the LNAs that are located well down within the feed structure. The 180 Ohm impedance of each of the two pairs of small cables provides an adequate match for each of the two 207 Ohm orthogonal polarizations. The magnitude of the power reflection from this mismatch is 0.5%. Although small, this mismatch is responsible for the fine frequency structure, ~0.5 GHz wide, in the final feed system temperature measurements.

There remain the inductances in the long leads that connect the coaxial cable centerlines to the tip terminals at the apex of the pyramid. To avoid the effect of this inductance, we absorb it by halving it for both leads of each polarization and adding a capacitance to ground at each mid point. By design, these inductances and added capacitances form low pass filters that have an upper band edge at 15 GHz for each polarization. The tip region, which is very compact, is shown on the right side of Figure (8). The circuit diagram for these structural parasitic components and their selected values are shown at the bottom of Figure (8); they provide a well matched Chebychev low pass filter (Zverev, 1967) for the connection between each feed polarization and its corresponding LNA at the end of each coaxial cable pair.

Also shown in the upper left of Figure (8) two of the four planar rexolite support structures that attach the feed arms to the central pyramid. Near the tip of the pyramid, the high frequency region, the standoff thicknesses are .25 mm, and near the large end, the low frequency region, they are 2.5 mm. Calculated reflections from these standoffs are less than -30 dB, and they have little detrimental effect on the operation of the feed.

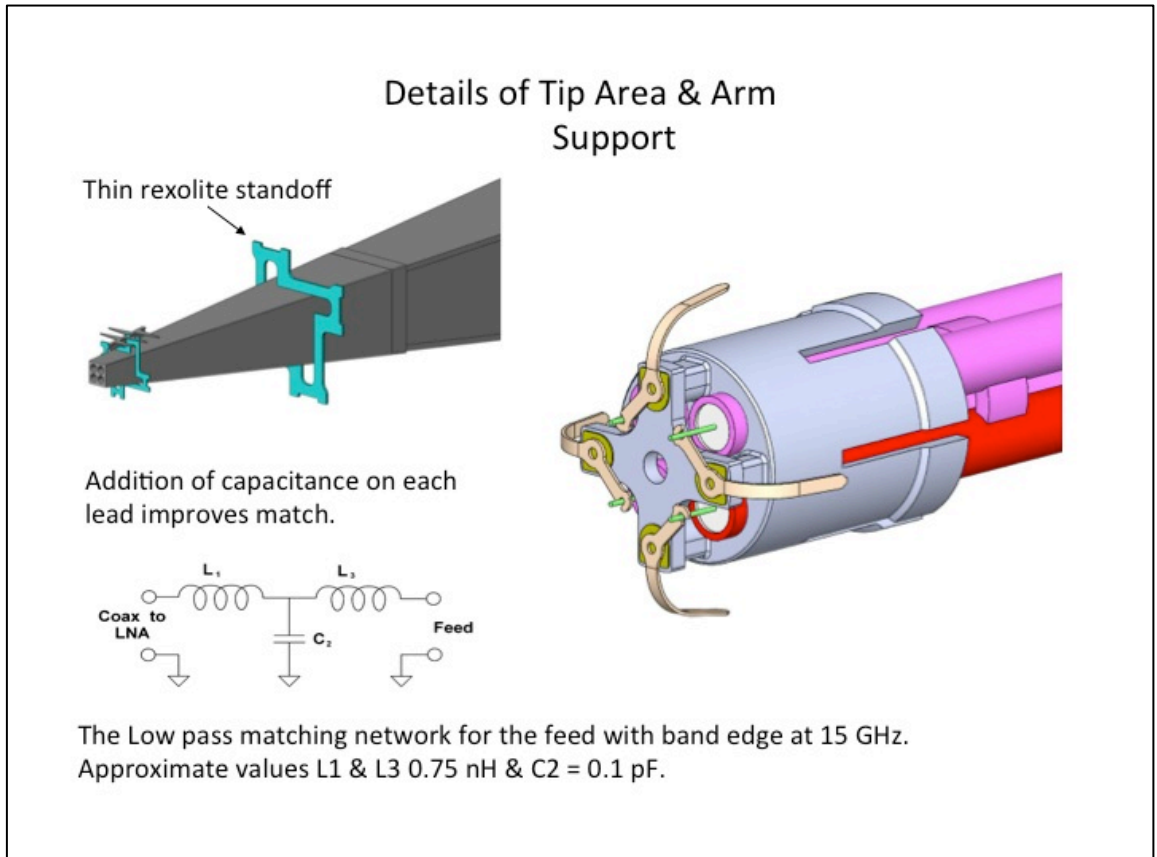

Figure (8) Details of Arm Support and the tip area for the Log Periodic Feed. Visible are the rexolite standoffs connecting the feed arms to the pyramid, details of the feed tip with coax cables, and capacitors that comprise the low pass filter circuit described (Zverev, 1967).

The efficacy of this low pass filter mounting at the tip is illustrated in Figure (9) that shows the calculated transmission of the low pass filter circuit model yielding an excellent result up to 14 GHz.

# Calculated Low Pass Filter Transmission Factor

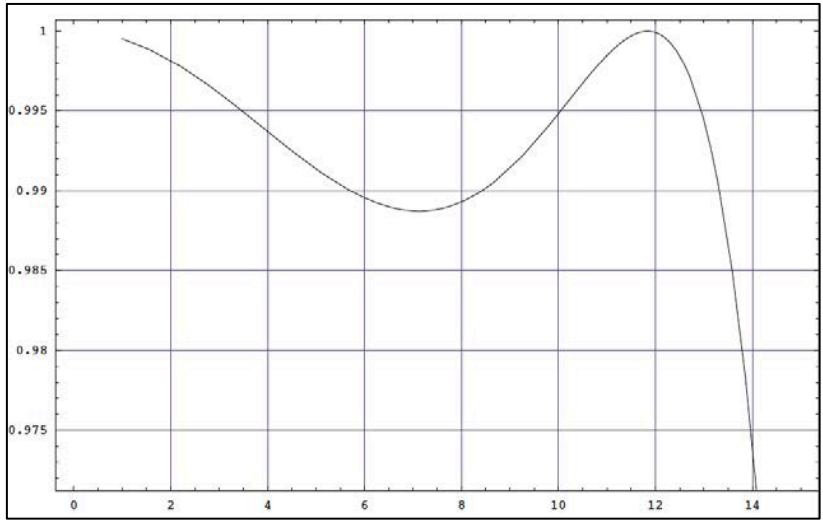

Figure (9)
Transmission factor in
the feed tip as a
function of frequency.

### Protection for Ultrawideband LNAs

In the presence of strong transmitters on aircraft, in orbit, and incorporated into groundbased electronic products, it is necessary to protect these ultrasensitive, wideband LNAs (LNF-ABLNC1_15A) from being damaged during observations.  Independent of the particular frequency band being selected by the observer, the LNAs are sensitive to strong signals at all frequencies within their frequency range, all the time.  The new Antonio feeds contain a pair of back to back Schottky diodes mounted to ground across each of the two input terminals as part of the custom mounting block for the LNA.  The diodes clamp when the voltage they experience exceeds 0.7 volts, corresponding to a power of 2.5 mW. These diodes are clearly visible in Figure (10). As an example; a typical avionics transponder used in general aviation aircraft flying through the antenna main beam, at an altitude of 1000 m, will produce a power flux of 0.02 mW/m², or a 0.4 mW peak pulse focused directly onto the LNA in an antenna. This power level will not engage the Schottky diodes, but a CAL FIRE helicopter flying over the array at much lower altitudes during fire season (a not infrequent event) would do so.  A live cell phone in the pocket of a member of a film crew, when the bomb bay doors of the shroud are opened for better viewing (also a not infrequent event), poses a threat as well.

### The Cryocooler

 It is essential to bring the physical temperature of the feed down to near 70 K so that its thermal noise contributes very little to the receiver noise of the system.  With the feed in the vacuum bottle, connecting the base of the feed to the same cryocooler that cools the LNA will work if it has adequate capacity.  The heat that must be removed is the general 300 K infrared background radiation around the bottle and the feed.  This radiation will transmit through the bottle readily and illuminate the feed.

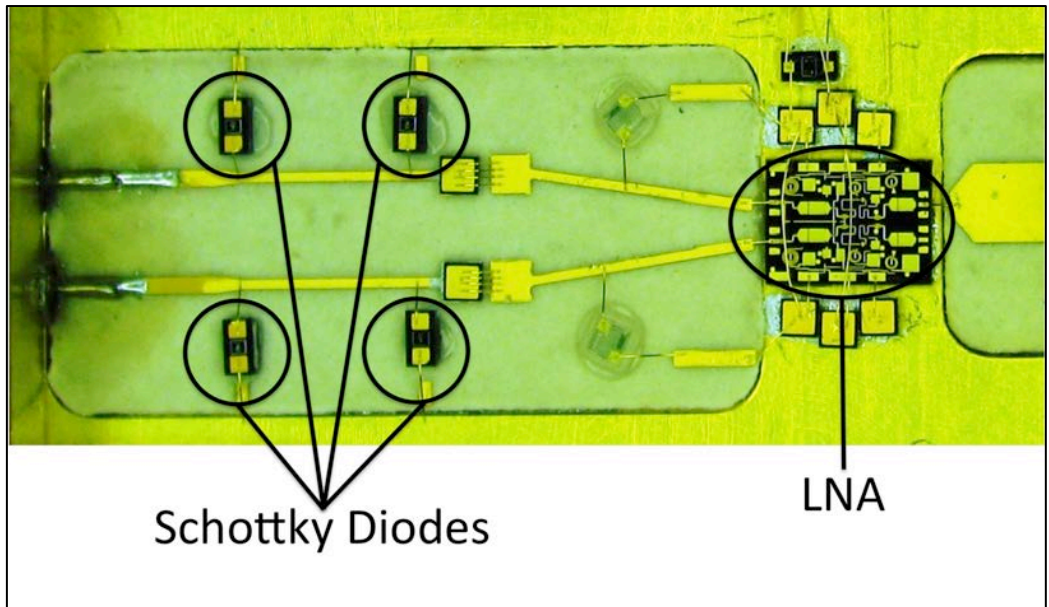

Figure (10) – A micrograph of a portion of the mounting block for the new amplifiers. Two pairs of back to back Schottky diodes combine with the inductances of the strip line segments to produce a wideband, low pass filter with an upper band edge at 15 GHz, to protect the LNF-ABLNC1_15A LNA. Image courtesy of Low Noise Factory.

It is practical to cool the entire feed in the dewar to a temperature to between 65 K and 70 K with a modest refrigerator connected just to its large end. The thermal conductivity of both the pyramid and the feed arms is adequate to maintain a small temperature difference between the feed tip and the refrigerator in the presence of the 300K infrared environment. The total infrared thermal load on the refrigerator is calculated to be about 2.5 Watts. The Sunpower CryoTel GT cooler was selected, and the temperature of the feed in operation is measured to be close to 70 K as expected.

During the 12 months following the installation of the first ten new feeds on ATA antennas, the physical temperatures of their LNAs have been measured daily. The LNA temperatures have remained steady and close to 65 K. These are important measurements because a reservoir of ambient cooling air is blown through buried plastic conduits from the node buildings and across the cryocoolers at each antenna, to keep them from overheating. In the summer, when the soil is very dry, the heat exchange between the colder ground and the warmer air is less efficient. Nevertheless, the performance of the cryocooler system has been adequate to maintain stable regulation of the physical temperature of the LNAs throughout every season.

Because the feed operates cantilevered from the primary dish, at the antenna focus, it is critical to control any structural vibrations within the unit. An additional vibration isolation mechanism was designed and built at Minex Engineering and integrated into the Sunpower CryoTel GT cryocooler (Fleming, private communication 2014). The residual peak to peak vibration measured at the tip of the feed is < 0.001".

## Measured Receiver System Noise Temperatures

The receiver system temperatures are measured with the receivers installed on the antennas.  The two reference loads used for these Y-factor measurements are the atmosphere and an ambient absorber that can be placed over the entire feed.  The ambient load is a metal cylinder with AEMI Broadband Convoluted absorber AEC-4 covering its inner walls and end plate.  This material shows a power reflection coefficient less than .001 and a one-way absorption greater than 25 dB at all frequencies in the range 0.9 to 15 GHz. The atmospheric brightness is based on an accurate model for the Hat Creek site that uses the known atmospheric properties and the 20° elevation angle at which the measurements are made. The azimuthal angles of all test measurements were chosen to avoid any background contribution from the Milky Way Galaxy at low frequencies. The ambient load temperature is measured with an accurate electronic thermometer, a typical value being 298 +/- 0.1 K.

Figure (11) shows the upgraded feed mounted inside the antenna, covered by a bag made of Sunbrella fabric, that provides additional safety in the event of a failure in the pyrex vacuum bottle.  Also shown is the ambient load that is placed around the entire feed during measurements of the system noise temperature to provide an excellent black body.

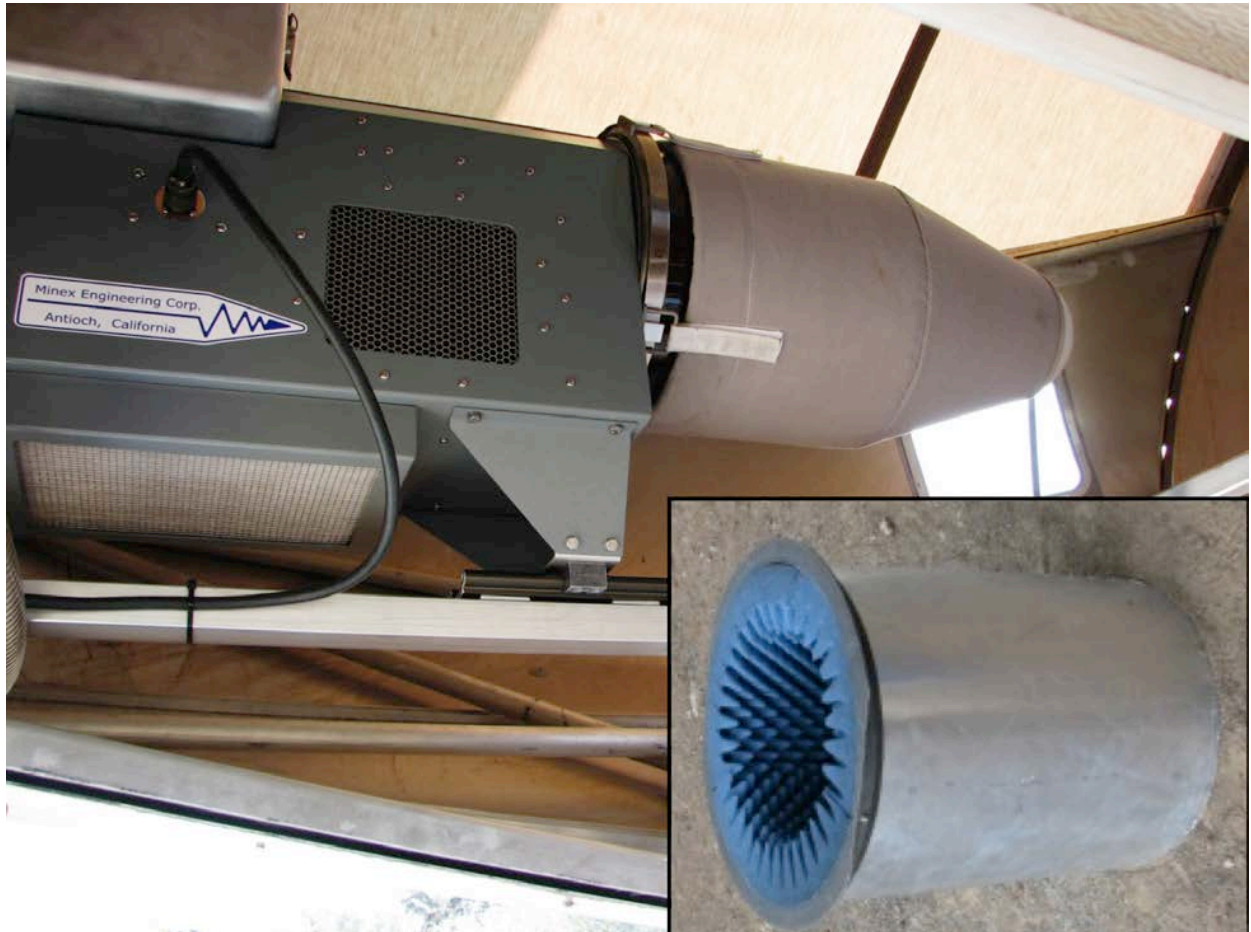

Figure (11)  Upgraded feed mounted on antenna.  Inset shows the hot load that surrounds the feed during measurement of system noise temperature.

A major design goal for these new feeds is to enable observation of the entire, quiet, terrestrial microwave window from 0.9 GHz to 14 GHz with as little added instrumental noise as possible.  Figure (12) shows how emission from the $O_2$ and $H_2O$ vapor in the atmosphere must add to the measured receiver temperatures in Figure (13) at the higher frequencies and synchrotron emission from the galaxy at lower frequencies as well. Measured system temperatures for both polarizations for ten feeds installed during October and December of 2015 appear in Figure (13).

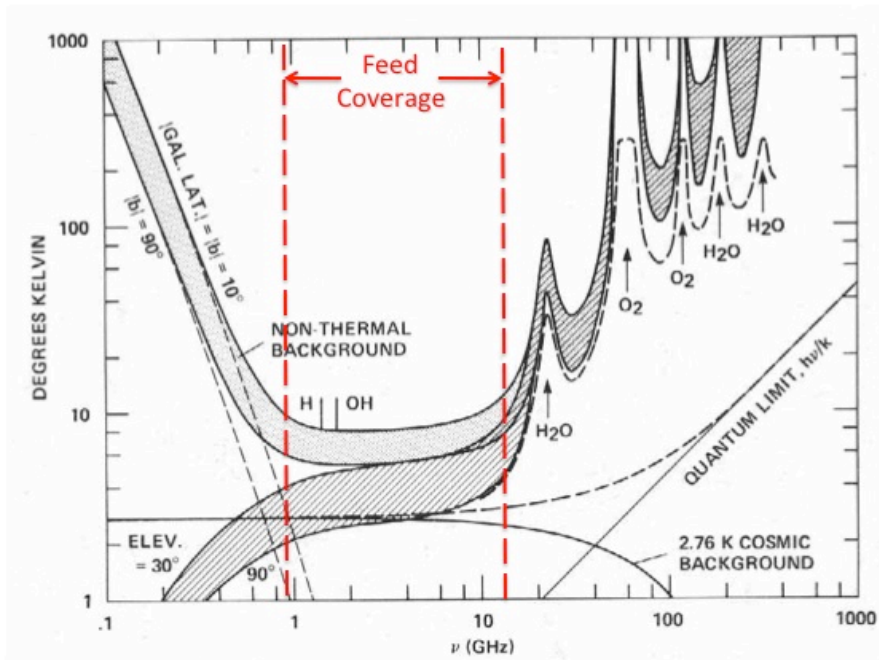

Figure (12) This plot shows the model radio brightness in the terrestrial microwave window
(NASA –SP419, 1976). The lowest background region is 1 – 15 GHz at a mean galactic latitude. The measured
ATA background in Figure (13) shows system temperatures slightly above these limiting values, as discussed
above.

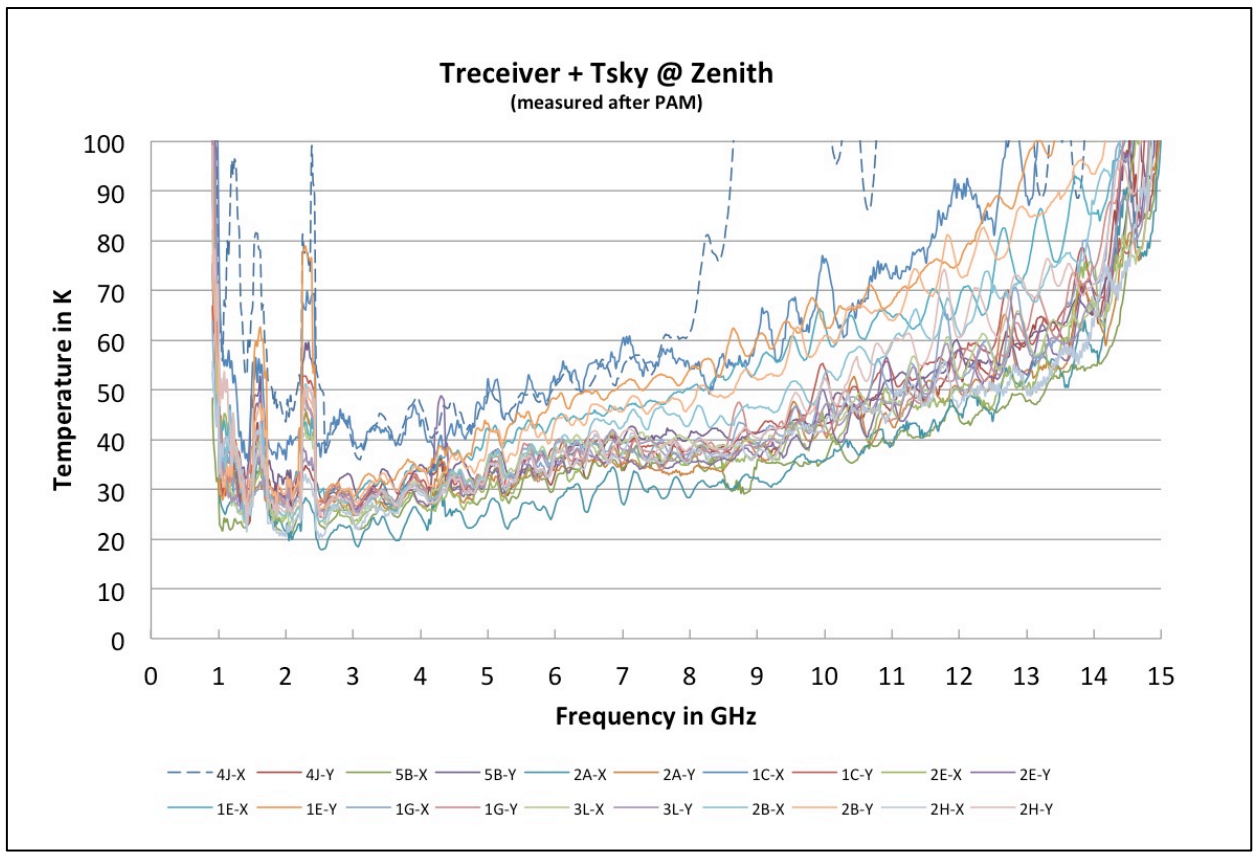

Figure (13) System temperatures measured at the ATA on ten new feeds during October and December of 2015, including both polarizations. An atmospheric model was used to correct the 20° elevation measurements to zenith. The spikes at the low frequencies are satellite transmissions (XM, Glonass, GPS, Inmarsat) and air traffic responses to ARTC from aircraft. The small vertical fluctuations that have a period of about 0.5 GHz are consistent with the small mismatch between the coaxial cables (180 ohms) and the feed (207 ohms) discussed above. The performance of 4J-X is unacceptable; that feed is scheduled for repair.

## Acknowledgements

It is a pleasure to acknowledge the philanthropic support from Franklin Antonio that enabled this upgrade. Research and manufacturing of the LNAs (LNF-ABLNC1_15A) used in this upgrade have been carried out at the GigaHertz Centre in a joint project financed by Swedish Governmental Agency of Innovation Systems (VINNOVA), Chalmers University of Technology, and Low Noise Factory, Omnisys Instruments, SP and Wasa Millimeter Wave. The SETI Institute provided scientific staff to analyze the measurement results and SRI International provided on-site staff to assist with the installation and physical measurements on the antennas.